\documentclass[12pt]{iopart}

\expandafter\let\csname equation*\endcsname\relax
 \expandafter\let\csname endequation*\endcsname\relax
\usepackage{amsmath}
\usepackage[utf8]{inputenc}
\usepackage{latexsym}
\usepackage{color}
\usepackage{graphicx}
\usepackage{epsfig}
\usepackage{amsopn}
\usepackage{amsfonts}
\usepackage{amssymb}
\usepackage{subfigure}
\usepackage{braket}
\usepackage{epstopdf}
\usepackage{braket}
\usepackage{nicefrac}
\usepackage[colorlinks=true,
pdfpagelabels,
pdfstartview = FitH,
bookmarksopen = false,
bookmarksnumbered = false,
linkcolor = black,
plainpages = false,
hypertexnames = false,
citecolor = black] {hyperref}

\begin{document}

\title{Fast thermometry for trapped ions using dark resonances}
\author{J Ro\ss nagel, K N Tolazzi, F Schmidt-Kaler, K Singer}
\address{QUANTUM, Institut f\"ur Physik, Universit\"at Mainz, D-55128 Mainz, Germany}
  
\ead{j.rossnagel@uni-mainz.de}
\begin{abstract}
We experimentally demonstrate a method to determine the temperature of trapped ions which is suitable for monitoring fast thermalization processes. We show that observing and analyzing the lineshape of dark resonances in the fluorescence spectrum provides a temperature measurement which accurate over a large dynamic range, applied to single ions and small ion crystals. Laser induced fluorescence is detected over a time of only $20\,\mu$s allowing for rapid determination of the ion temperature. In the measurement range of $10^{-1}-10^{+2}\,$mK we reach better than $15\,\%$ accuracy. Tuning the cooling laser to selected resonance features allows for controlling the ion temperatures between $0.7\,$mK and more than $10\,$mK. Experimental work is supported by a solution of the 8-level optical Bloch equations when including the ions classical motion. This technique paves the way for many experiments comprising heat transport in ion strings, heat engines, non-equilibrium thermodynamics or thermometry of large ion crystals.
\end{abstract}

\maketitle
\tableofcontents

\section{Introduction}
Controlling and measuring the temperature of trapped ions is essential for many different applications. Especially in elementary quantum information processors \cite{cirac1995quantum,blatt2008entangled,haffner2008quantum,hanneke2009realization} and for quantum simulations with linear ion crystals \cite{friedenauer2008simulating,kim2010quantum,lanyon2011universal,blatt2012quantum} 
the initialization at very low temperatures near the motional ground state is crucial. On the other hand, there is growing interest in the field of quantum thermodynamics and techniques like reservoir engineering or dissipative state preparation have moved into focus of ion trap experiments \cite{berut2012experimental,toyoda2013experimental,porras2004bose,lin2013dissipative,kienzler2015quantum}. 
One of the major quests here is to prepare and measure thermal states in an intermediate range between a few and up to some hundreds of phonons. For this range, sideband spectroscopy and ion manipulations with sequences of coherent laser pulses are no longer ideal tools to control or measure the motional state.

We have adapted and implemented a measurement scheme which is based on the multi-level electronic structure and the occurrence of dark resonances in the fluorescence signal of trapped ions. 
These narrow dark resonance features due to coherent population trapping in a three level system~\cite{alzetta1976experimental,arimondo1976nonabsorbing,arimondo1996v} provide a high velocity sensitivity of their fluorescence signal. Dark resonances have been shown to provide a useful tool for compensating micromotion in ion traps or tuning the cooling rate for single ions~\cite{lisowski2005dark,reiss2002raman,eschner2003laser}. 
Recently they have been used to determine the temperatures of a laser cooled neutral atom cloud in the sub-millikelvin regime~\cite{peters2012thermometry}. 
The technique demonstrated here for measuring temperatures is tailored for applications in thermodynamical machines, for large ion crystals, and to observe fast thermal processes, which typically occur in the range of $10^{-4}\,$s.
It can be applied far outside the Lamb-Dicke regime. While dark resonance thermometry can be adopted within a wide range of temperatures it is convenient to use and does not require any additional lasers or technical components as compared to Doppler cooling. 

This new method complements the thermometry toolbox for trapped ions at an intermediate range, while for highly excited thermal states there are several well established methods to determine the temperature, such as Doppler recooling, well-suited for a temperatures range 
$10^{-1}-10^{1}\,$ meV ($10^3-10^5\,$mK)~\cite{epstein2007simplified,wesenberg2007fluorescence,huber2008transport,daniilidis2011fabrication}, or spatial thermometry \cite{knunz2012sub,walther2012precision}. For low thermal states close to the motional ground state, precise results can be achieved by sideband spectroscopy \cite{epstein2007simplified,turchette2000heating,brownnutt2014ion}, where the motional state is deduced from the relation of excitation probabilities on the sidebands of a narrow transition. The temperature range can be expanded using specially shaped pulse sequences \cite{walther2012controlling}. However, in experiments with low trap frequencies motional sidebands can not be resolved. For those setups dark resonance thermometry now enables to measure low temperatures in the range of $10^{-1} - 10^{+2}\,$mK .

We organized the paper such that we start with a derivation of the appearance of dark resonances in a multi-level system from the steady-state solution of the optical Bloch equations. We analyze their influence on the temperature of the ions in a semi-classical picture. Then, we present experimental results on the measurement of the steady state temperature, followed by a section where we show how dark resonances can be employed for tuning the ion temperature. Finally, we present measurements of the dynamics of heating and cooling processes.

\section{Theoretical aspects of the dark state thermometry with trapped ions}

\begin{figure}
\centering
\includegraphics[width=0.8\textwidth]{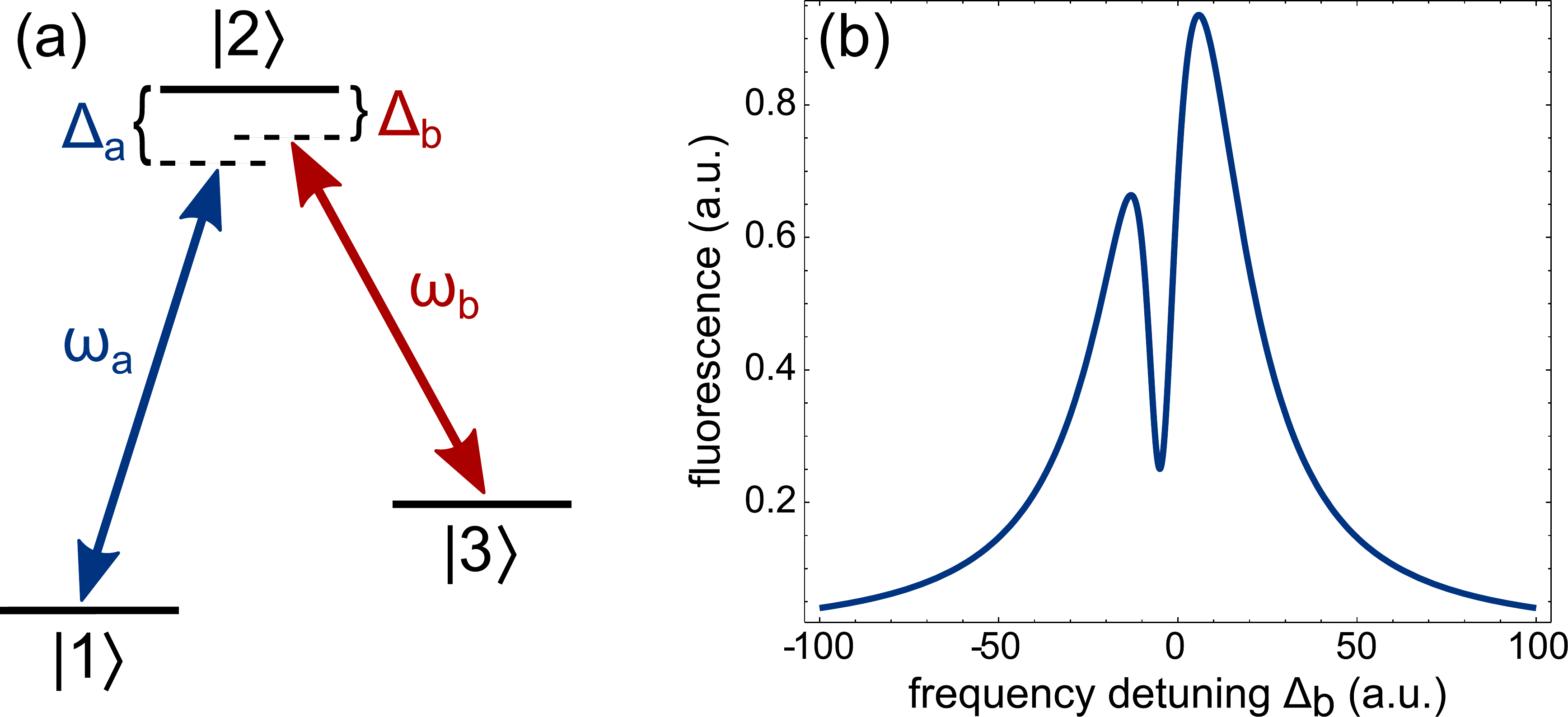}
\caption{(a) Three-level system with atomic states $\ket{1}$, $\ket{2}$ and $\ket{3}$, and two near-resonant light fields with frequencies $\omega_{a}$ and $\omega_b$ in $\Lambda$-configuration. (b) To obtain a  spectrum of resonance fluorescence the detuning $\Delta_a$ is scanned. If both detuning are equal $\Delta_a = \Delta_b$ the fluorescence shows a drop, the dark resonance.} 
\label{fig:3-level-CPT}
\end{figure}

As simplest case, we consider an atomic three-level system, e.g. a $\Lambda$-configuration, with states $\ket{1}$, $\ket{2}$, $\ket{3}$, where two near-resonant light fields with frequencies $\omega_{a}$ and $\omega_b$ drive the atomic transitions $\omega_{12}$ and $\omega_{23}$, respectively (see Fig.~\ref{fig:3-level-CPT}(a)). In the case that the detunings $\Delta_i$ of both light fields are chosen such as $\Delta_a-\Delta_b=0$ a coherent superposition of the states $\ket{1}$ and $\ket{3}$ is generated. If the radiation lifetime of $\ket{1}$ and $\ket{3}$ is much longer than the lifetime of $\ket{2}$ coherent population trapping is achieved, resulting in a dark state decoupled from the driving fields \cite{cohen2011advances}. The steady state of the internal dynamics is generally reached much faster than any relevant external timescales, such as coherent oscillations or changes in temperature.
When scanning the frequency of one of the two fields while keeping the other constant, a drop in the fluorescence spectrum of $\ket{2}\rightarrow \ket{1}$ is observed (see Fig.~\ref{fig:3-level-CPT}(b)). 

For an ion in motion, this dark resonance line is modified due to the Doppler effect. As a thermal state of an ion is described by many different velocity classes, this results in a broadening of the dark resonance. Additional broadening is caused by the phase fluctuations of the  laser fields, denoted by a spectral linewidth $\Gamma$. However, if the spectral linewidth of the lasers is known, the broadening of the dark resonance feature allows to determine the temperature of the ion by comparing with calculated spectra.

\subsection{Theoretical description of dark resonances with ions in motion}

For a quantitative determination of the ion temperature from the measured fluorescence spectra, a model for the theoretical lineshape is required. Therefore we describe the dynamics of the system by a Lindblad master equation, the time evolution of the density matrix. We consider the Hamiltonian of the system 
\begin{equation}
\mathcal{H}=\mathcal{H}_{dip}+\mathcal{H}_{ion}
\label{eq:}
\end{equation}
which is composed of $\mathcal{H}_{dip}$ the ion-light dipole interaction and
$\mathcal{H}_{ion}$,  describing internal states of the ion including the detunings of the two light fields $\Delta_a$ and $\Delta_b$~\cite{cohen2011advances,scully1997quantum,fleischhauer2005electromagnetically}. 
The motional state of the ion, not included in the Hamiltonian, will be introduced later as a spectral Doppler shift of the detunings $\Delta_{a,b}$.  

The Lindblad master equation for the density matrix $\hat{\rho}$ reads \cite{cirac1992laser}
\begin{equation}
\frac{d \hat{\rho}}{dt}= - \frac{i}{\hbar} \left[ \mathcal{H},\hat{\rho} \right] + \mathcal{L}(\hat{\rho}).
\label{eq:Lindblad}
\end{equation}
The spontaneous decay of the population of the excited levels is introduced by the dissipative Lindblad operator
\begin{equation}
\mathcal{L}(\hat{\rho})=\sum_i{\left[ \mathcal{C}_i\hat{\rho} \mathcal{C}^\dagger_i-\frac{1}{2} \left( \mathcal{C}_i^\dagger \mathcal{C}_i \hat{\rho} + \hat{\rho} \mathcal{C}^\dagger_i \mathcal{C}_i \right) \right]}.
\label{eq:LindbladCoupling}
\end{equation}
where the transition operators $\mathcal{C}_i=\sqrt{\Gamma_{pq}} |q\rangle\langle p|$ describe the dissipation along the dipole transitions $p \rightarrow q$. They depend on the decay rates according to Fermi's golden rule \cite{scully1997quantum,cohen1975atoms}
\begin{equation}
\Gamma_{ab}=\frac{8 \pi^2}{3\epsilon_0 \hbar \lambda_{pq}} \left| \langle p| \vec{D} |q \rangle \right|^2
\label{eq:}
\end{equation}
with the atomic dipole operator $\vec{D}$ and the transition wavelength $\lambda_{pq}$.

The spectral widths of the light fields $\Gamma_a$ and $\Gamma_b$ are included as an additional broadening of the corresponding atomic level $\ket{q}$. Phase fluctuations between the atomic state and the light field lead to a decay of the dark state and are modeled by transition operators $C_{a,b}=\sqrt{\Gamma_{a,b}} \ket{q}\bra{q}$~\cite{cohen1975atoms,cohen2011advances} (see appendix Eq.~\ref{eq:LaserLinewidth1} and \ref{eq:LaserLinewidth2}). 

To get the resulting spectrum of the fluorescence light the steady state solution $d\hat{\rho}/dt = 0$ is required. In the following we will solve this equation numerically. We transform Eq.~(\ref{eq:Lindblad}) into a matrix equation, in which the density matrix $\hat{\rho}$ with dimension $N \times N$ is represented by a vector with $N^2$ entries, where $N$ is the number of atomic levels involved. Consequently, the master equation with the Lindblad operator $L_{ik}$ in matrix form is described by
\begin{equation}
\frac{d\rho_i}{dt}=\sum_{k=1}^{N^2}{L_{ik} \rho_k} =0 \quad \mathrm{with} \quad i\in \left\{ 1,...,N^2 \right\}.
\label{eq:matrix}
\end{equation}
One of these equations (for simplicity $i=1$) is replaced by the normalization condition $\mathrm{Tr}(\hat{\rho})=1$. Finally, to obtain the steady state solution $\rho^0$ we solve equation (\ref{eq:matrix}) by numerical matrix inversion. The fluorescence rate $\mathcal{F}$ is proportional to the population in the excited state(s). To get a fluorescence spectrum including the dark resonances, as shown in Fig.~\ref{fig:3-level-CPT}(b), $\mathcal{F}$ has to be calculated for every detuning $\Delta_i$ of the corresponding light field within the range of interest.

\begin{figure}[tbp]
\centering
\includegraphics[width=\textwidth]{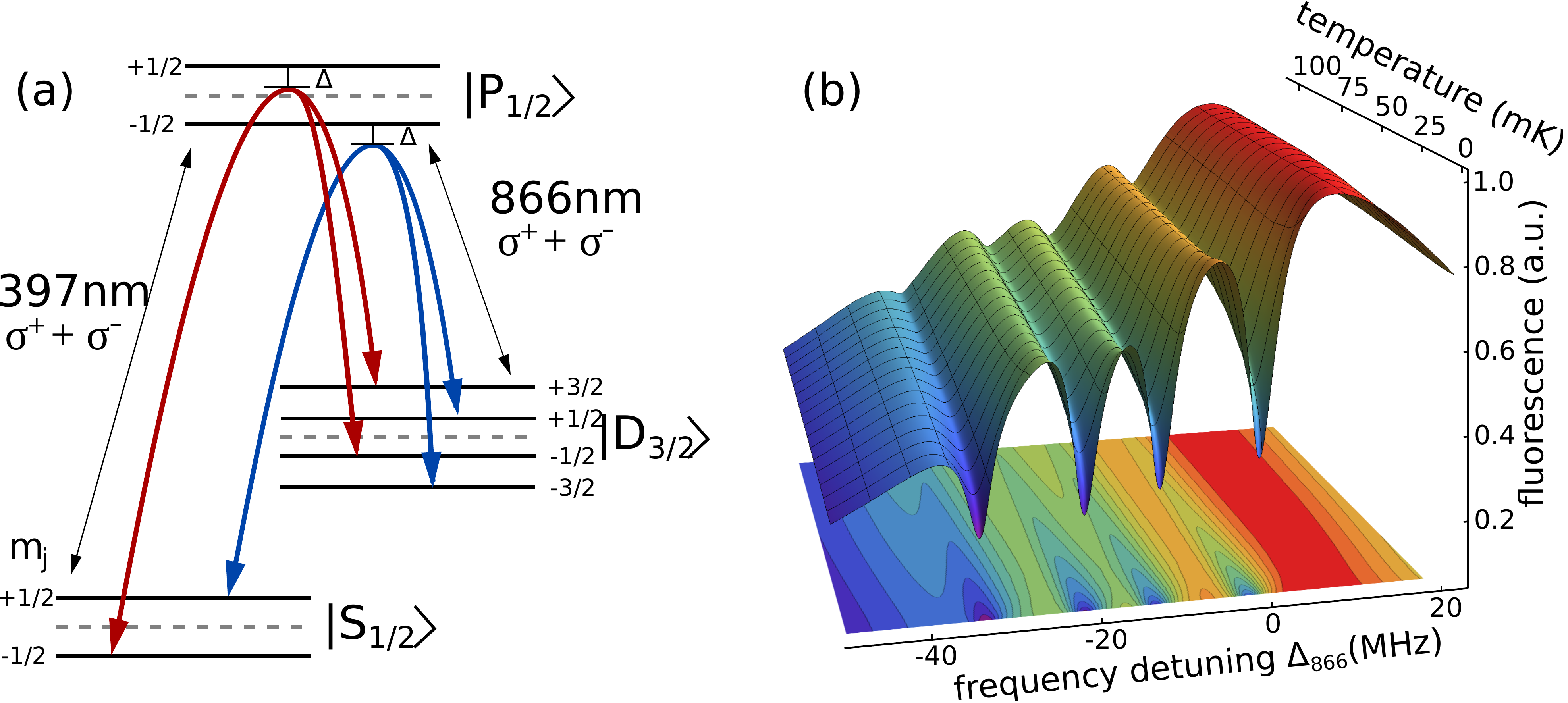}
\caption{(a) Zeeman splitting lifts the degeneracy of the three-level system (dashed lines), leading to a 8-level system, not drawn to scale. The polarization of the two light fields leads to four possible coherent transitions from the ground states $^2S_{1/2}$ into the metastable states $^2D_{3/2}$. (b) The corresponding fluorescence spectrum is calculated as a function of the frequency detuning $\Delta_{866}$ and as a function of the temperature of the ions for collinear laser beams. The decreasing depths of the resonances for increasing temperatures is emphasized by the contour plot on the bottom, where blue~(red) indicates low (high) fluorescence.}
\label{fig:levelScheme}
\end{figure}

Now, we take the finite temperature of the ion into account, assuming a thermal state at temperature~$T$.  For conditions with phonon numbers $n\gg 1$, the motional state of the ion can be introduced in a semi-classical way. We consider the two light fields to have an angle~$\alpha$ with respect to each other. 
A velocity $\vec{v}$ of the ion results in a Doppler shift of the two frequencies $\omega_a$ and $\omega_b$ and thus shifts their detuning by $\delta\Delta_i(v,\alpha) = \vec{k}_i \cdot \vec{v}$, where $\vec{k}_{i}$ are the corresponding wave-vectors. Through the inclusion of the Doppler shifted detunings in the calculation of the spectrum we get a fluorescence rate $\mathcal{F}(\alpha,\vec{v})$ depending on $\alpha$ and $\vec{v}$. We obtain a temperature-dependent fluorescence rate $\widetilde{\mathcal{F}}(T)$
\begin{equation}
\widetilde{\mathcal{F}}(T)=\int{ P_T(\vec{v})\cdot \mathcal{F}_\alpha(\vec{v}) \;d\vec{v}},\label{eq:integral}
\end{equation}
 by integrating over all velocity classes  in the Maxwell-Boltzmann distribution
\begin{equation}
P_T(\vec{v})=\left( \frac{m}{2\pi k_B T} \right)^{3/2} e^{ \left( -m\vec{v}\,^2/2 k_B T\right) }
\label{eq:}
\end{equation}
with atomic mass $m$ and Boltzmann-constant $k_B$. For a numerical solution of Eq.~(\ref{eq:integral}) we discretize $\vec{v}$ into distinct velocity classes along the directions of the laser beams. 
The line shape of the dark resonances as a function of the temperature $T$ is shown in Fig.~\ref{fig:levelScheme}(b). To determine the ion temperature the measured fluorescence data is fitted by $\widetilde{\mathcal{F}}(T)$ using a Markov chain Monte-Carlo parameter estimation \cite{poschi_private}.

For $^{40}$Ca$^+$ ions with an electronic ground-state configuration of [Ar]$\,4\,^2S_{1/2}$, having one single $S$ electron above a closed shell, the energetic structure can be approximated by a three-level system with fine structure levels $^2S_{1/2} =\ket{1}$, $^2D_{3/2}=\ket{3}$ and $^2P_{1/2}=\ket{2}$, see Fig.~\ref{fig:levelScheme}(a). 
The $S_{1/2} \leftrightarrow P_{1/2}$ dipole transition is used for laser cooling and is driven by a laser near~$397\,$nm. 
In order to reach a closed cycling transition all the population which decays into the metastable state $D_{3/2}$ with a lifetime of $1.2\,$s is pumped back to the $P_{1/2}$ state by a laser near~$866\,$nm. In our experimental setting both laser frequencies are red-detuned with respect to the corresponding atomic transition $\Delta_{397,866} < 0$. 
To obtain a resonance fluorescence spectrum, scattered photons near $397\,$nm are recorded as a function of the detuning $\Delta_{866}$, while $\Delta_{397}$ is kept constant. Note, that scanning $\Delta_{397}$ is unfavorable for thermometry, as its detuning has a strong influence on the cooling rate. 
As a magnetic field is applied to define the quantization axis, the degeneracy of the electronic states is lifted leading to an 8-level system \cite{schubert1995transient,oberst}, see Fig.~\ref{fig:levelScheme}(a). Depending on the polarizations of the laser beams up to 12 dark resonance lines can be observed. 
In our experiment the polarizations of both beams are chosen such that only $\sigma^+$ and $\sigma^-$ transitions are addressed, leading to the appearance of 4 dark resonances (Fig.~\ref{fig:levelScheme}). Regarding the $P_{1/2} \leftrightarrow D_{3/2}$ transition this polarization is avoiding dark states in the $D_{3/2}$ manifold. 
For the $S_{1/2} \leftrightarrow P_{1/2}$ transition a $\sigma^+ + \sigma^-$-polarization is convenient as it leads to an excitation of the transitions which are maximally apart and therefore most easily resolved.

The corresponding matrix form of $\mathcal{H}$ as well as the coupling operators $\mathcal{C}_i$ are given in the appendix. The Zeeman-splitting leads to a 64-dimensional Lindblad operator $L_{ik}$ in Eq.~(\ref{eq:matrix}) and thus to a time consuming matrix decomposition. To reduce the computational complexity we approximate the effect of temperature on the dark resonances by a relative Doppler broadening $\Gamma_D(T,\alpha)$. It can be deduced from the variance of the relative Doppler shift $\delta\Delta=\delta\Delta_{397}-\delta\Delta_{866}$ and reads
\begin{equation}
\Gamma_D(T,\alpha) = \left| \vec{k}_{397} - \vec{k}_{866} \right| \sqrt{\frac{k_B T}{2 m}} = \sqrt{ k^2_{397} +  k_{866}^2  -2\,k_{397} k_{866} \cos{(\alpha)}} \,\sqrt{\frac{k_B T}{2 m}}.
\label{eq:k-vectors}
\end{equation} 
Regarding the shape of the dark resonances, the laser linewidths and the Doppler broadening act in a similar way. Therefore, $\Gamma_D$ is treated as an additional contribution to the broadening due to the spectral widths of the lasers. During the following applications both methods, using Eq.~(\ref{eq:integral}) and Eq.~(\ref{eq:k-vectors}), have been compared for different spectra and lead to consistent results for the temperatures within their uncertainties. The thermal broadening $\Gamma_D(T,\alpha)$ in Eq.~(\ref{eq:k-vectors}) shows a strong dependence on the relative orientation $\alpha$ of the $k$-vectors of the two lasers. The strongest influence of temperature, and thus the highest sensitivity even at low temperatures, is achieved for counter propagating laser beams ($\alpha=\pi$), where $\Delta_{397}$ and $\Delta_{866}$ are Doppler shifted with opposite signs. Already for moderate temperatures of a few millikelvin the contrast of the dark resonances becomes too small to assign a reliable temperature. In this range, however, the best choice is a collinear beam configuration ($\alpha=0$), which has reduced sensitivity, but is well suited for a wide range of temperatures up to several tens of millikelvin, as shown in Fig.~\ref{fig:levelScheme}(b). 

\subsection{Shaping the laser cooling by exploiting dark resonances}\label{sec:bath}
The modification of the photon scattering rate on the laser cooling transition in vicinity of dark resonances affects the equilibrium ion temperature. In the following this effect is analyzed in a semi-classical picture, where the Doppler shift reduces the light scattering of selected velocity classes.

For Doppler cooling, during which the $866\,$nm laser may be far detuned from dark resonances, the cooling rate is dominated by the spontaneous emission lifetime $\tau=7\,$ns for the decay of $P_{1/2}$ into $S_{1/2}$ \cite{jin1993precision,sahoo2009relativistic,safronova2011blackbody}. 
The period of the secular motion is much larger than the lifetime, $\omega \ll 1/\tau$ (weak binding regime) and hence the ion does not move considerately during the scattering process \cite{stenholm1986semiclassical,metcalf1999laser}. 
The steady state of the internal dynamics is typically reached after $10^{-7}\,$s and therefore, in our case, much faster than any secular motion.
In vicinity of a dark resonance the velocity dependent Doppler cooling force is affected by the modified scattering rate. To analyze its influence we introduce the effective Doppler shift $\delta\Delta^\textrm{DR}$ of the relative frequency detuning of the laser frequencies with respect to a dark resonance $\Delta^\textrm{DR}=\Delta_{866} - \Delta_{397}$. Including $\delta\Delta^\textrm{DR}$ the condition for a dark resonance reads
\begin{equation}
\Delta^\textrm{DR}_\delta=\Delta^\textrm{DR}-\delta\Delta^\textrm{DR}=\Delta_{866}-\Delta_{397} - \left(\nu_{866}-\nu_{397}\right)\frac{\vec{v}\cdot \vec{e}}{c} =0,
\label{eq:}
\end{equation}
where $\nu_i$ are the absolute frequencies of the lasers, $\vec{e}$ is their direction and $\vec{v}$ the velocity vector of the ion. The relative Doppler shift $\delta\Delta^\textrm{DR}$ has the opposite sign of the two individual Doppler shifts, because $\nu_{866} < \nu_{397}$. If the ion moves in the same direction as the laser beams $\vec{v} \uparrow \uparrow \vec{k}_i$, photon scattering leads to an amplification of the motion of the ion. 
This heating process limits the achievable temperature during Doppler cooling~\cite{metcalf1999laser}. 
However, if the frequency of the $866\,$nm laser is red detuned with respect to the dark resonance ($\Delta^\textrm{DR}<0$), those velocity classes with $\vec{v} \uparrow \uparrow \vec{k}_i$  are Doppler shifted into resonance $\Delta^\textrm{DR}_\delta \approx 0$. 
Photon scattering then is suppressed and the resulting excitation of the ion motion is reduced. As a consequence, for $\Delta^\textrm{DR}<0$ a lower temperature can be achieved, similar to the velocity dependent coherent population trapping (VSCPT) schemes for samples of neutral atoms \cite{aspect1988laser}.

\begin{figure}
\centering
\includegraphics[width=0.65\textwidth]{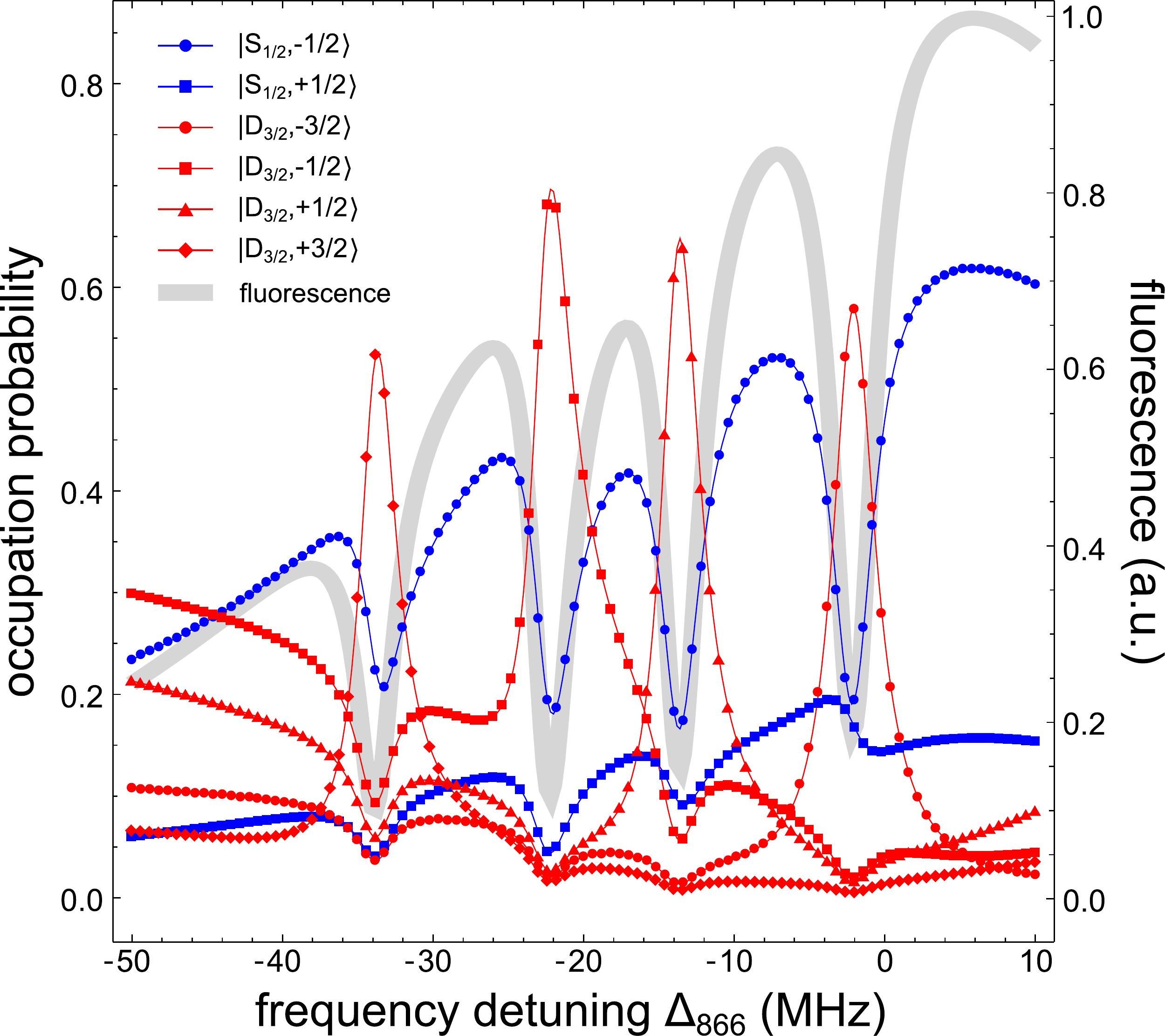}
\caption{The calculated occupation probabilities $\hat{\rho}_{ii}$ for Zeeman sublevels of the $S_{1/2}$ (blue lines) and $D_{3/2}$ manifold (red lines), symbols serve to distinguish between different Zeeman states. The resonance fluorescence (gray) is proportional to the occupation in $\ket{P_{1/2}}$. Near a dark resonance, the maximum occupation changes from the $S_{1/2}$ manifold (blue line) to the corresponding $D_{3/2}$ sublevel (red line).}
\label{fig:occupation}
\end{figure}

In a complementing analysis we show that scattering on motional-sidebands of the coherent two photon $\Lambda$-transition $S_{1/2} \leftrightarrow D_{3/2}$ is enhanced near to dark resonances. In Fig.~\ref{fig:occupation} we present the result of a numerical analysis of the populations of the individual $S_{1/2}$ and $D_{3/2}$ sublevels in vicinity of dark resonances. On resonance with the $P_{1/2} \leftrightarrow D_{3/2}$ transition, far detuned from any dark resonances with $\Delta^\textrm{DR} \ll 0$, the decay from $P_{1/2}$ to $S_{1/2}$ is favored over the decay from $P_{1/2}$ to $D_{3/2}$ and therefore the population is accumulated in the $S_{1/2}$ states \cite{gerritsma2008precision,ramm2013precision}. However, this situation is inverted if $\Delta^\textrm{DR} \approx 0$. As can be seen from Fig.~\ref{fig:occupation} the population in the $D_{3/2}$ manifold then is higher than in the $S_{1/2}$ states. For this reason the direction of the $\Lambda$-transition from $D_{3/2}$ to $S_{1/2}$ is favored over the reverse direction. As a consequence, for a negative detuning $\Delta^\textrm{DR} < 0$ the effective transition from $D_{3/2}$ to $S_{1/2}$ removes phonons from the motional mode, while phonons are added for $\Delta^\textrm{DR}>0$ \cite{reiss2002raman}. The cooling cycle is closed by an excitation into the $P_{1/2}$ states and subsequent spontaneous decay. In the steady-state solution each $S_{1/2}$ sublevel is effectively coupled to all resonances because a strong $S_{1/2} \leftrightarrow P_{1/2}$ scattering ensures a constant redistribution of the population within the $S_{1/2}$ manifold. Therefore, the population of both $S_{1/2}$ levels shows four resonances. The different shape of the population of the two $S_{1/2}$ sublevels is due to the detuning of the $397\,$nm laser, which leads to different scattering rates.

\section{Experimental realization of thermometry with small ion crystals}

\begin{figure}[tbp]
\centering
\includegraphics[width=0.8\textwidth]{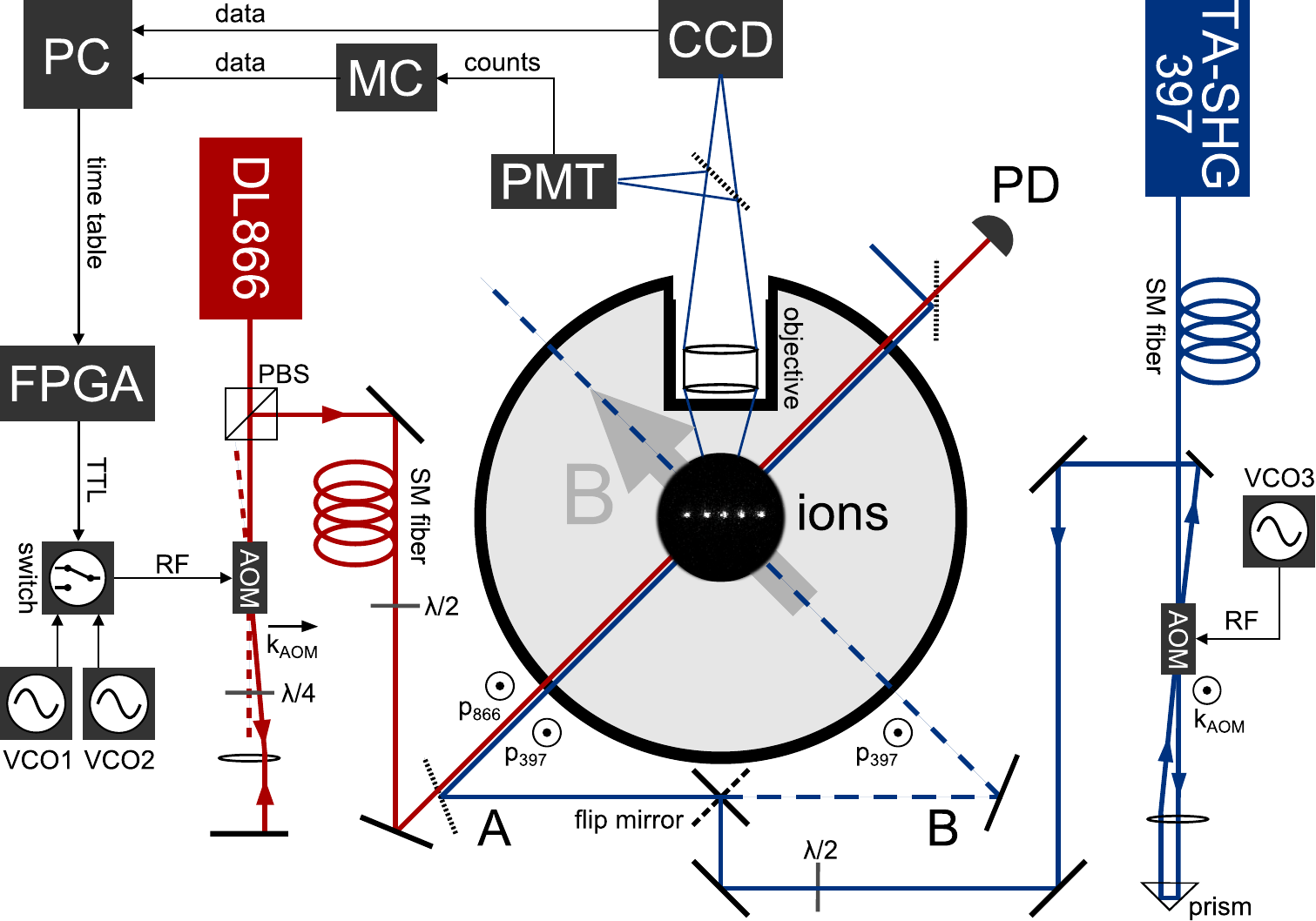}
\caption{Experimental setup with laser and imaging system. A flip mirror allows the $397\,$nm laser beam to illuminate the ions in parallel (A) or perpendicular (B) configuration with respect to the $866\,$nm laser beam. AOMs in double pass configuration  control the intensity and tune the frequency of both laser beams. Switching the radio frequency of the AOM in the beam path of the $866\,$nm laser between two voltage controlled oscillators allows for short pulses, controlled by the computer (PC) and a field-programmable gate-array (FPGA). The polarization $p_i$ of both beams is chosen perpendicular to the magnetic field $\vec{B}$ in order to excite both $\sigma^+$ and $\sigma^-$ transitions. An objective system images the fluorescence light at $397\,$nm onto a CCD camera as well as onto a photo-multiplier tube (PMT), whose counts are recorded by a micro-controller (MC). }
\label{fig:setup}
\end{figure}

\subsection{Setup of the experiment}\label{Sec:setup}
The experiments are performed with $^{40}$Ca$^+$-ions, trapped in a linear Paul trap with four radial electrodes of a diameter of $0.8\,$mm and a distance to the trap axis of $1.5\,$mm. Each two diagonally opposing electrodes are driven by the same phase of a radio-frequency at $\omega_{RF}/2\pi=21.45\,$MHz, but with a  shift of $\pi$ with respect to the other two electrodes. A peak to peak voltage of about $900\,$V leads to radial frequencies $\omega_{x}/2\pi$ and $\omega_{y}/2\pi$ of $468\,$kHz and $472\,$kHz, respectively. Applying a constant voltage of $9\,$V to the end-caps at a distance of $8\,$mm to each other, results in an axial trap frequency of $\omega_{ax}/2\pi = 170\,$kHz. The $^{40}$Ca$^+$-ions are loaded into the trap by ionizing neutral calcium atoms effusing from a thermal oven using a resonant two step photoionization process. The ions are Doppler cooled on the $S_{1/2} \leftrightarrow P_{1/2}$ transition. A static magnetic field of $4.7\times10^{-4}\,$T is applied with $45^\circ$ to the trap axis, which lifts the degeneracy of the fine structure levels. The laser system consists of two extended-cavity diode-lasers. One amplified (Tapered Amplifier) and frequency doubled (Second Harmonic Generation) diode-laser provides laser-light at $397\,$nm (\textit{Toptica} TA-SHG 110) and a second diode-laser at $866\,$nm (\textit{Toptica} DL100). A sketch of the full experimental setup is shown in Fig.~\ref{fig:setup}. Both of the lasers are locked to an external reference cavity using the Pound-Drever-Hall technique. From the shape of the in-loop Pound-Drever-Hall error signal we have deduced the full linewidths at half maximum of the two lasers to $450\pm 30\,$kHz for the $397\,$nm laser and $490 \pm 30\,$kHz for the $866\,$nm laser. Acousto-optical modulators (AOM) are installed in the beam paths of both lasers using a double pass configuration to switch the beams on and off, as well as to allow for frequency shifts. A field programmable gate array (FPGA) controls an rf-switch to change between two different rf-drive frequencies of the AOM in the branch of the $866\,$nm laser and thus allows for fast experimental sequences with nanosecond precision. For the spectral scans a constant intensity of the $866\,$nm laser beam over a range of $40\,$MHz is required. To avoid power fluctuations due to a frequency dependent diffraction efficiency of the AOM (\textit{Brimrose} TEF-270-100), the beam intensity is measured on a photodiode (PD) as a function of the AOM drive frequency and its driving power is corrected accordingly.

The two laser beams are oriented parallel to one another and perpendicular to the magnetic field (beam path \textit{A} in Fig.~\ref{fig:setup}). They have an angle of $45^\circ$ to the axis of the trap and $60^\circ$ to the radial directions, and thus all vibrational modes are addressed simultaneously. The polarization of both laser beams is chosen by waveplates to be perpendicular to the magnetic field as well, leading to a superposition of $\sigma^+$ and $\sigma^-$ polarization with respect to quantization axis. The intensities of both laser beams are $35\,\mu$W at a beam waist of $160\,\mu$m for the $397\,$nm laser and $80\,\mu$W at a beam waist of $380\,\mu$m. A flip-mirror permits to change the beam path of the $397\, $nm laser to propagate parallel to the direction of the magnetic field (beam path \textit{B}, dashed line in Fig.~\ref{fig:setup}). The polarization of this beam path is again chosen by waveplates such that both $\sigma^+$ and $\sigma^-$ transitions are driven. For beam path \textit{B} the orientation of the two laser beams is perpendicular to one another and the Doppler shift does not cancel to first order anymore. Consequently, this configuration leads to a higher sensitivity on the ion movement and therefore is more sensitive to temperature differences. 

The fluorescence light at $397\,$nm is collected by an objective lens system with a numerical aperture of $0.26$. The image of the trapped ions is projected onto a CCD camera. A pickup mirror reflects $30\,\%$ of the collected photons onto a photomultiplier tube (PMT). The camera image is used for trapping, controlling and preparing the ions. During the experimental sequences we use the PMT to be able to count single photons, which enables us to measure with exposure times of some microseconds. The PMT events are counted by a $40\,$MHz micro-controller (MC), which is triggered by the FPGA to guarantee precise timing with respect to the laser pulses.

\subsection{Experimental procedures}\label{sec:procedure}

To obtain a spectrum of resonance fluorescence the frequency of the $866\,$nm beam is scanned via the drive frequency of the according AOM. We can chose two different modes of operation. 
First, a wide scan of the spectrum including all dark resonances is recorded, where the frequency is tuned continuously over the regions of interest and scattered photons near 397~nm are recorded. 
A fit of the data by the calculated fluorescence rate $\mathcal{F}(\Delta_{866})$ allows to determine Rabi-frequencies $\Omega_{397}$ and $\Omega_{866}$, detuning of the $397\,$nm laser $\Delta_{397}$, magnitude of the magnetic field $B$. 

Then, to greatly reduce the influence of scattered photons on the temperature of the ion we use an alternative mode of data acquisition, which consists of 2 steps: (i)~For an equilibration time $t_\textrm e \geq 2\,$ms the frequency of the $866\,$nm beam is set at some fixed detuning $\Delta_e$, resulting in a steady-state of the ion temperature. (ii)~Then, for a much shorter measurement time of typically $t_\textrm m = 20\,\mu$s, the drive frequency of the AOM is switched to the detuning $\Delta_m$. Only during $t_m$ scattered fluorescence photons are recorded on the PMT. The sequence of steps (i) and (ii) is repeated multiple times for detecting in total about $10^4$ photons to reduce the statistical error. Then, $\Delta_m$ is stepped to the next value to obtain the full shape of one dark resonance for determining the temperature $T$. All other parameters, such as B-field, or Rabi frequencies had been determined from the scan over the full spectrum. Unless noted otherwise, the collinear beam configuration (beam path A) is employed to make use of a maximum range of temperatures which are accessible.

\subsection{Engineering thermal states}
\begin{figure}[tbp]
\centering
\includegraphics[width=\textwidth]{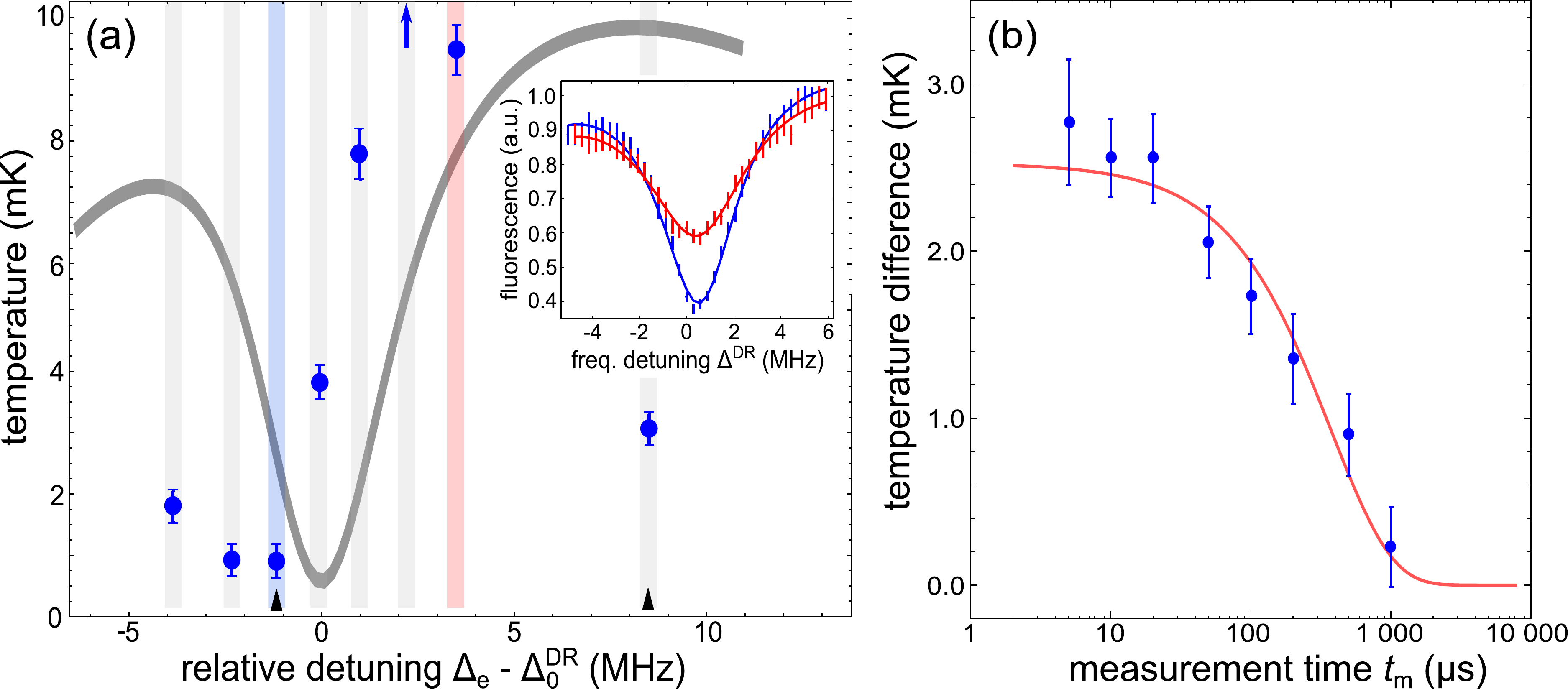}
\caption{(a) Measured equilibrium temperatures (blue) while varying the detuning $\Delta_\textrm e$. Temperatures between $0.7(4)\,$mK and $9.4(6)\,$mK are determined from the shape of the dark resonance, see inset.  At $\Delta^\textrm{DR}_0 + 2\,$MHz the measured temperature diverges. The gray line indicates the relative position within the resonance feature. (b)~Influence of the measurement time $t_\textrm m$ on the resulting temperature. The difference of the measured temperatures for $\Delta_\textrm e=\Delta^\textrm{DR}_0 + 8.5\,$MHz and $\Delta_\textrm e=\Delta^\textrm{DR}_0 - 1.2\,$MHz (triangular markers in (a)) has been evaluated as a function of the measurement time $t_\textrm m$. An exponential fit reveals a time constant of $\tau_\textrm m=0.38(7)\,$ms.}
\label{fig:bath}
\end{figure}

The multi-level system of $^{40}$Ca$^+$-ions allows to tailor the line shape and vary the resulting temperature. Here, we demonstrate how the temperature can be controlled and measured in vicinity of the dark resonance at $\Delta^\textrm{DR}_0$ corresponding to the $\ket{D_{3/2}, -3/2}$  state, see Fig.~\ref{fig:bath}. We apply the measurement procedure where the frequency of the $866\,$nm laser is switched, as described above. If $\Delta_\textrm e$ is close to the $P_{1/2} \leftrightarrow D_{3/2}$ transition and far from any dark resonances during the equilibration $t_\textrm e$ in step (i), we measure a temperature of  $3.1(4)\,$mK. However, if the frequency of the $866\,$nm laser is close to a dark resonance but red detuned $\Delta_\textrm e<\Delta^\textrm{DR}_0$ during $t_\textrm e$, we see that the temperature drops by a factor of four and measure $0.7(4)\,$mK at $\Delta_\textrm e = \Delta^\textrm{DR}_0 - 1.2\,$MHz. In contrast, we observe strong heating effects if $\Delta_\textrm e$ is blue detuned to the dark resonance $\Delta_\textrm e>\Delta^\textrm{DR}_0$. At $\Delta_\textrm e = \Delta^\textrm{DR}_0+2\,$MHz the ion is excited so much that the dark resonance feature, recorded during step (ii), vanishes completely. We can only give a lower bound of $T>80\,$mK.

Because different frequencies along the dark resonance lead to different equilibrium temperatures, we investigate experimentally the influence of the measurement duration $t_\textrm m$ in step (ii) on the resulting temperature. Therefore, we vary $t_\textrm m$ between a few~$\mu$s and $1\,$ms. In the latter case, $t_\textrm m$ is sufficiently long to let the thermal state equilibrate to a temperature determined by $\Delta_\textrm m$ during the measurement. For short times of $t_\textrm m$, however, the temperature is governed by $\Delta_\textrm e$ and not significantly influenced by the measurement. To characterize the influence of the measurement we have tuned $\Delta_\textrm e$ to two different values at $\Delta^\textrm{DR}_0 -1.2\,$MHz and $\Delta^\textrm{DR}_0 +8.5\,$MHz (triangular markers in Fig.~\ref{fig:bath}(a)), corresponding to temperatures of $0.7(4)\,$mK and $3.1(4)\,$mK, respectively. The equilibration process is shown in Fig.~\ref{fig:bath}(b), where the measured temperature difference $\Delta T$ is plotted as a function of $t_\textrm m$. An exponential fit of the temperature differences by $\Delta T(t_\textrm m)=a\,\exp(-t_\textrm m/\tau_\textrm m)$ reveals a decay time of $\tau_\textrm m=0.38(7)\,$ms, corresponding to a rate of about 2.6~kHz. Indeed, for measurement times $t_\textrm m \leq 20\,\mu$s the modification of the measured temperature of $\Delta T \leq 0.1\,$mK is neglectable as compared to the concurrent errors. At low temperatures the dominant error results from the systematic uncertainty of the spectral linewidth $\Gamma$ of the laser sources. An uncertainty in the knowledge of the laser linewidth of $30\,$kHz results in a temperature uncertainty of $0.4\,$mK. 

\subsection{Extending the range of stationary temperatures}
\begin{figure}[tbp]
\centering
\includegraphics[width=0.8\textwidth]{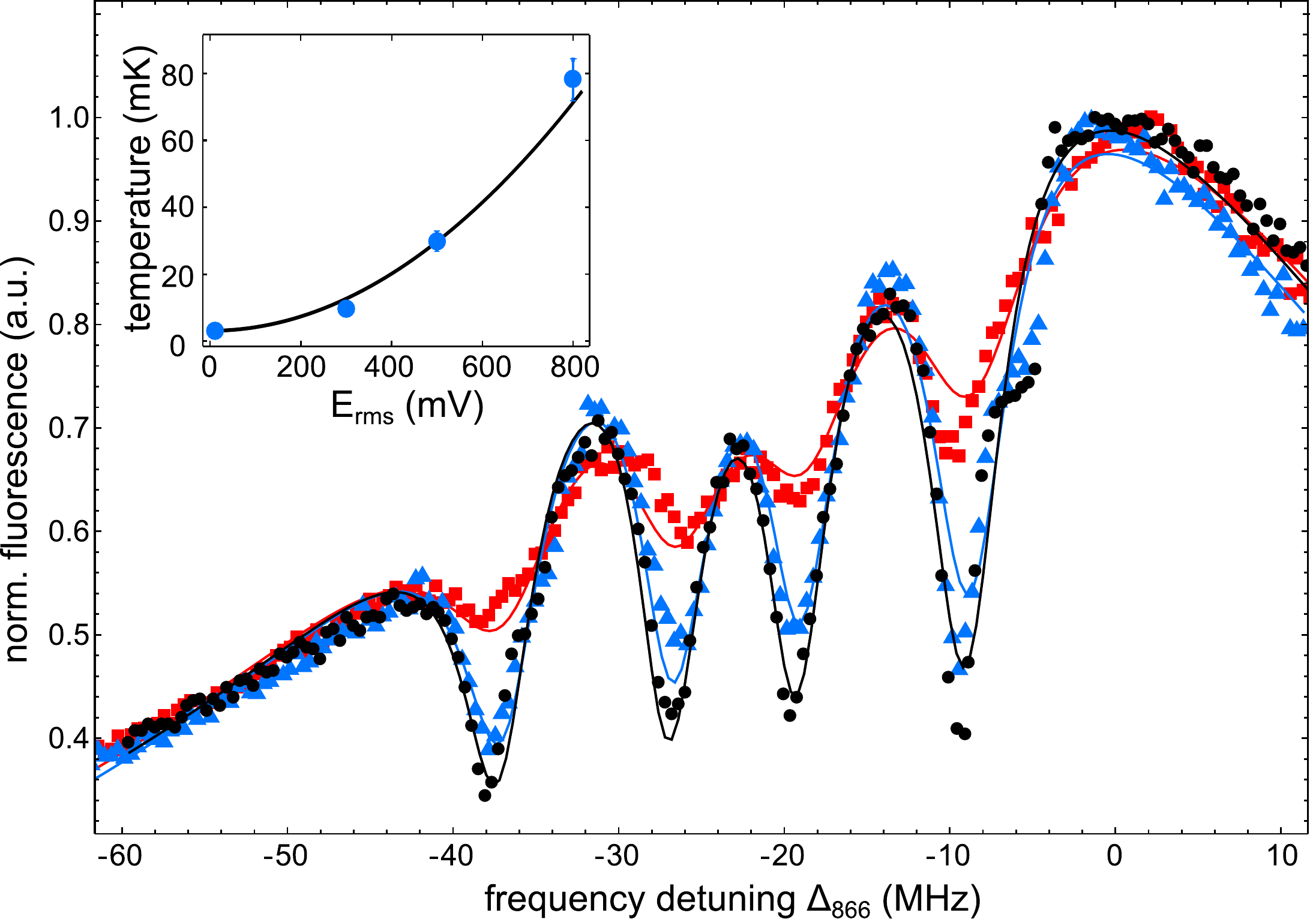}
\caption{Dark resonances for different applied electronic noise amplitudes ($0\,$V$_{\textrm{rms}}$~(black), $300\,$mV$_{\textrm{rms}}$~(blue), $800\,$mV$_{\textrm{rms}}$~(red)). A fit yields $\Omega_{397}/2\pi=12\,$MHz, $\Omega_{866}/2\pi=8\,$MHz, $\Delta_{397}/2\pi=14\,$MHz and $B=4.7\times10^{-4}\,$T. The effective temperatures result to $3.1(5)\,$mK (black), $9(1)\,$mK (blue) and $46(4)\,$mK (red). The inset shows the measured temperature as a function of the noise amplitude, following~$T \propto E _\textrm {rms}^2$.}
\label{fig:thermal}
\end{figure}

For many applications, it might be important to extend the range of control also to thermal states with higher temperature. To this end we apply electrical noise to the trap electrodes, which are used for compensating micro-motion. This leads to a random displacement of the trap potential, while the trap frequency is constant. Combined with the dissipative process of laser cooling it is applied for $t_\textrm e\geq 2\,$ms, such that a steady-state with adjustable temperature $T$ is achieved, where $T$ is controlled by the amplitude of the electric field noise as $T \propto E _\textrm {rms}^2$. This environment mimics an effective thermal heat bath for the trapped ion \cite{cirac1994quantum,poyatos1996quantum}. We perform the measurements with a linear crystal of 5 ions and could vary its temperatures over more than one order of magnitude. Fig.~\ref{fig:thermal} shows the spectrum at different noise amplitudes. We fit temperatures of $T= 3.1(5)\,$mK, $9(1)\,$mK and $46(4)\,$mK to the data. Note, that for these scans, we have used the continuous acquisition scheme. At high noise amplitudes, where the dark resonances are broadened, we find excellent fit to the model curves. For the data without excitation we observe deviations from the fit curve near $\Delta_{866}=-6\,$MHz which are explained by cooling/heating in vicinity of the dark resonance, an artifact absent in the pulsed acquisition mode.

\subsection{Investigation of thermal dynamics}

\begin{figure}[tbp]
\centering
\includegraphics[width=0.9\textwidth]{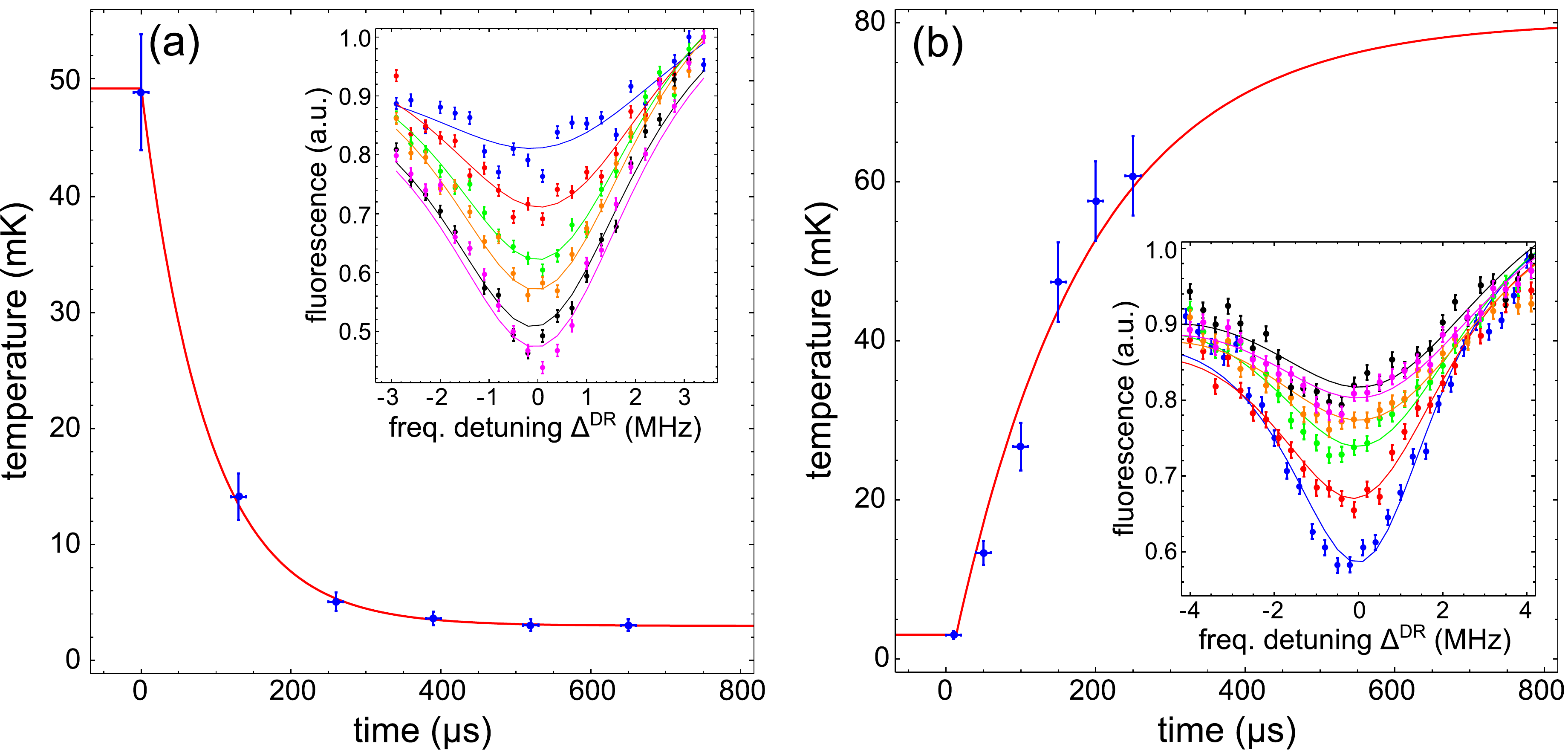}
\caption{(a) Cooling of an ion crystal: Ions are simultaneously excited by electrical noise and cooled ($\Delta_\textrm e= 0$) resulting in a temperature of $T_\textrm i=71(7)\,$mK as measured from the shape of dark resonances, see inset. Once the electrical noise is switched off, the laser cooling reduces $T$ with an exponential $\tau_\textrm c=87(1)\,\mu$s reaching finally $T=3.0(5)\,$mK. (b) Ion heating dynamics: Starting with laser a cooled crystal, electrical noise increases $T$ with $\tau_\textrm h=257(40)\,\mu$s a final $T=71(7)\,$mK. In both cases we use short exposure times of $t_\textrm m=20\,\mu$s.}
\label{fig:heatingCooling}
\end{figure}

To capture the time evolution of temperatures we apply an electrical noise pulse of $2\,$ms while laser cooling the ions with a fixed detuning $\Delta_{866}$ to reach an initial steady state temperature $T_i$. 
Then we switch the noise off and observe the thermal re-cooling process. Therefore we record fluorescence spectra by the pulsed measurement technique with different time delays between $50$ and $650\,\mu$s after the noise pulse, see Fig~\ref{fig:heatingCooling}(a). Fitting the measured temperatures by $T \propto \exp{(-\Delta t/\tau_\textrm c)}$ reveals a time constant of $\tau_\textrm c=87(3)\,\mu$s. Similarly, we observe the inverse heating process: Starting with a laser cooled ion we measure the re-equilibration process after switching on the noise pulse, see Fig.~\ref{fig:heatingCooling}(b). An exponential fit results in a heating time of $\tau_\textrm h=257(40)\,\mu$s. In the language of reservoir engineering \cite{cirac1994quantum,poyatos1996quantum}, such  dynamics may be interpreted as thermalization of the ion with an effective heat bath. Once the noise pulse is switched the ion is out of equilibrium and subsequent re-equilibration can be observed.

\section{Outlook and applications}

We have shown that dark resonances in a multi-level electronic system, e.g. in $^{40}$Ca$^+$, are a versatile tool for controlling and measuring the temperature of a single ion or small ion crystal. We observe and interpret the modification of the dark resonances and determine the temperature of the ions from a fit of the laser-induced fluorescence data by the model of the lineshape. This technique is convenient to use as it only requires the light sources which are used for Doppler cooling and to excite the ion's resonance fluorescence. Using this type of thermometry we have been able to generate and measure temperatures between $0.7\,$mK and $61\,$mK.
We see advantages of dark resonance thermometry especially for applications with low trap frequencies, as well as in cases where not all modes of motion of an ion crystal are cooled to the ground state of vibration and thus resolved sideband techniques are not applicable. Therefore, our method fills the gap between the sideband spectroscopy technique, used for very low temperatures close to the motional ground state, and the Doppler recooling method, which can be applied to hot thermal states with temperatures above $10^3\,$mK. Note that the multi-level dynamics lead to Doppler recooling results which do require extended numerical studies for extracting a temperature.
This new developed and tested method would be well suited for studies of non-linear dynamics in ion crystals. Non-linear couplings may be studied in systems with many ions and in the presence of thermal excitations with two-dimensional spectroscopy \cite{lemmer2014two}, in order to detect signatures of a structural phase transition of the ion crystal, as well as resonant energy exchange between modes. For the investigation of structural phase transitions in ion crystals and the formation of defects \cite{ulm2013observation} it is important to determine and measure the temperature. Moreover, we have demonstrated that a measurement of thermal dynamics can be performed on the timescale of a few microseconds. This will allow for measuring thermal equilibration processes in other non-equilibrium situations. Having such a fast and almost non-invasive method to determine the temperature is an important prerequisite for future thermodynamic experiments in ion traps, such as time-resolved transport of thermal energy through an ion crystal \cite{bermudez2013controlling,lin2011equilibration,pruttivarasin2011trapped}. 
By the combination of electrical noise and laser cooling an effective and adjustable heat bath for an ion can be mimicked. This technique will be employed for driving  thermodynamic cycles of a single ion heat engine \cite{abah2012single,rossnagel2014nanoscale}, converting a temperature difference of two heat baths into coherent ion motion which equals mechanical work. As the frequency of the proposed thermodynamic cycle will be about $50\,$kHz, this requires a temperature manipulation and a temperature measurement technique which probes faster than $20\,\mu$s and still covers a large range of temperatures. Dark resonance thermometry will meet this requirements for characterizing the efficiency of the single ion engine. 

\ack
 We thank U.~G.~Poschinger for programming the fitting routine and for valuable discussions, as well as C.~Champenois and R.~Gerritsma for carefully reading the manuscript.  Furthermore, we acknowledge support by the Volkswagen-Stiftung, the DFG (Contract No. LU1382/
4-1) and the COST action MP 1209 “Thermodynamics in the quantum regime”.
 
\newpage
\section*{References}
\bibliography{literature}

\newpage

\section{Appendix: Remarks on the numerical solution of the lineshape}\label{App.1}

To calculate the Hamiltonian $\mathcal{H}$ of the 8-level system, we have set the zero point of energy to the $^2 P_{1/2}$ fine structure level and have transformed the light field Hamiltonian to the rotating frame of reference \cite{cohen2011advances,scully1997quantum,oberst}. Note that due the polarization of the laser beams only $\sigma^+$ and $\sigma^-$ transitions are driven.
The full matrix representation of $\mathcal{H}$ reads as follows:
\begin{small}
\begin{equation}
\begin{bmatrix}
-\frac{1}{2}g_S B-\Delta_{a} &0 &0            &-\frac{1}{\sqrt{3}} \Omega_{a}    &0               &0             &0            &0\\
    0            &\frac{1}{2}g_S B-\Delta_{a}&-\frac{1}{\sqrt{3}}\Omega_{a}   &0             &0               &0             &0            &0\\
    0            &-\frac{1}{\sqrt{3}}\Omega_{a}   &-\frac{1}{2}g_P B   &0             &-\frac{1}{2} \Omega_{b}      &0             &\frac{1}{2\sqrt{3}}\Omega_{b}&0\\
    -\frac{1}{\sqrt{3}}\Omega_{a}   &0            &0            &\frac{1}{2}g_P B     &0               &-\frac{1}{2\sqrt{3}}\Omega_{b}&0            &\frac{1}{2}\Omega_{b}\\
    0            &0            &-\frac{1}{2} \Omega_{b}   &0             &-\frac{3}{2}g_D B-\Delta_{b}&0             &0            &0\\
    0            &0            &0            &-\frac{1}{2\sqrt{3}}\Omega_{b}&0               &-\frac{1}{2}g_D B -\Delta_{b}&0            &0\\
    0            &0            &\frac{1}{2\sqrt{3}}\Omega_{b}&0             &0               &0             &\frac{1}{2}g_D B-\Delta_{b}&0\\
    0            &0            &0            &\frac{1}{2} \Omega_{b}     &0               &0             &0            &\frac{3}{2}g_D B -\Delta_{b}
\end{bmatrix}
\label{eq:HamiltonMatrix}
\end{equation}
\end{small}

with Land\'e-factors $g_S =2$, $g_P = \frac{2}{3}$, $g_D = \frac{4}{5}$ and the Zeeman energy in units of frequency $B=\frac{\mu_B|\vec{B}|}{\hbar}$. Further the Hamiltonian depends on the detunings of the lasers from their corresponding transition frequency $\Delta_a=\omega_{SP}-\omega_{397}$ and $\Delta_b=\omega_{PD}-\omega_{866}$ as well as the corresponding Rabi frequencies $\Omega_{a,b}$.

The coupling operators $\mathcal{C}_i$ from equation (\ref{eq:LindbladCoupling}) are composed of the decay rates $\Gamma_{pq}$ of the Zeeman sublevels $p\leftrightarrow q$, taking into account the Zeeman coherences \cite{oberst,cohen1975atoms}. This leads to six individual decay channels from the states $|P,m_j\rangle$ to $|S,m_j\rangle$ and $|D,m_j\rangle$ with the magnetic quantum number $m_j$
\begin{equation}
\Gamma_{PS}=\frac{8 \pi^2}{3\epsilon_0 \hbar \lambda_{SP}} \sum_{m=-\nicefrac{1}{2}}^{+\nicefrac{1}{2}}{\left| \langle P,+\tfrac{1}{2}| \vec{D} |S,m \rangle \right|^2} = \frac{8 \pi^2}{3\epsilon_0 \hbar \lambda_{SP}}  \sum_{m=-\nicefrac{1}{2}}^{+\nicefrac{1}{2}}{\left| \langle P,-\tfrac{1}{2}| \vec{D} |S,m \rangle \right|^2}
\label{eq:}
\end{equation}
and
\begin{equation}
\Gamma_{PD}=\frac{8 \pi^2}{3\epsilon_0 \hbar \lambda_{DP}} \sum_{m=-\nicefrac{3}{2}}^{+\nicefrac{3}{2}}{\left| \langle P,+\tfrac{1}{2}| \vec{D} |D,m \rangle \right|^2} = \frac{8 \pi^2}{3\epsilon_0 \hbar \lambda_{DP}} \sum_{m=-\nicefrac{3}{2}}^{+\nicefrac{3}{2}}{\left| \langle P,-\tfrac{1}{2}| \vec{D} |D,m \rangle \right|^2}.
\label{eq:}
\end{equation}

This allows us to calculate the transition operators
\begin{equation}
\mathcal{C}_i=\sqrt{\Gamma_{pq}}|q\rangle\langle p|
\label{eq:}
\end{equation}
for all non-zero dipole transitions. If we divide the transitions according to their polarizations, we find the following set of 6 transition operators:

\begin{align}
 &P_{1/2}\rightarrow S_{1/2}
 \begin{cases}\label{eq:COperatoren}
  \mathcal{C}_1=\sqrt{\frac{2}{3}\Gamma_{\text{P,S}}}\ket{S,-\tfrac{1}{2}}\bra{P,+\tfrac{1}{2}}\\\\
  \mathcal{C}_2=\sqrt{\frac{2}{3}\Gamma_{\text{P,S}}}\ket{S,+\tfrac{1}{2}}\bra{P,-\tfrac{1}{2}}\\\\
  \mathcal{C}_3=\sqrt{\frac{1}{3}\Gamma_{\text{P,S}}}\left(\ket{S,-\tfrac{1}{2}}\bra{P,-\tfrac{1}{2}}-\ket{S,+\tfrac{1}{2}}\bra{P,+\tfrac{1}{2}}\right)
 \end{cases}
 \\\nonumber\\
 &P_{1/2}\rightarrow D_{3/2}
  \begin{cases}
  \mathcal{C}_4=\sqrt{\frac{1}{2}\Gamma_{\text{P,D}}}\ket{D,-\tfrac{3}{2}}\bra{P,-\tfrac{1}{2}}+\sqrt{\frac{1}{6}\Gamma_{\text{P,D}}}\ket{D,-\tfrac{1}{2}}\bra{P,+\tfrac{1}{2}}\\\\
  \mathcal{C}_5=\sqrt{\frac{1}{6}\Gamma_{\text{P,D}}}\ket{D,+\tfrac{1}{2}}\bra{P,-\tfrac{1}{2}}+\sqrt{\frac{1}{2}\Gamma_{\text{P,D}}}\ket{D,+\tfrac{3}{2}}\bra{P,+\tfrac{1}{2}}\\\\
  \mathcal{C}_6=\sqrt{\frac{1}{3}\Gamma_{\text{P,D}}}\left(\ket{D,-\tfrac{1}{2}}\bra{P,-\tfrac{1}{2}}+\ket{D,+\tfrac{1}{2}}\bra{P,+\tfrac{1}{2}}\right)
  \end{cases}
 \end{align}

The linewidth of the lasers is introduced by an additional broadening of the corresponding $S_{1/2}$ and $D_{3/2}$ sublevels \cite{cohen1975atoms,oberst}
 \begin{align}\label{eq:LaserLinewidth1}
    \mathcal{C}_7&=\sqrt{\Gamma_{397}}\sum_{m=-\nicefrac{1}{2}}^{+\nicefrac{1}{2}}{\ket{S,m}\bra{S,m} }\\
    \mathcal{C}_8&=\sqrt{\Gamma_{866}}\sum_{m=-\nicefrac{3}{2}}^{+\nicefrac{3}{2}}{\ket{D,m}\bra{D,m} \label{eq:LaserLinewidth2}}
 \end{align}

\end{document}